\documentclass[twocolumn,showpacs,preprintnumbers,amsmath,amssymb]{revtex4}

\usepackage{graphicx}
\usepackage{dcolumn}
\usepackage{bm}
\usepackage{natbib}

\begin{document}

\title
{
Infrared spectroscopy under multi-extreme conditions: Direct observation of pseudo gap formation and collapse in CeSb
}

\author{T. Nishi$^{1}$}
\author{S. Kimura$^{1,2}$}\thanks{Electric address: kimura@ims.ac.jp}
\author{T. Takahashi$^{3}$}
\author{Y. Mori$^{4}$}
\author{Y.S. Kwon$^{2,5}$}
\author{H.J. Im$^{1}$}
\author{H. Kitazawa$^{6}$}

\affiliation{
$^1$Department of Structural Molecular Science, The Graduate University for Advanced Studies, Okazaki 444-8585, Japan\\
$^2$UVSOR Facility, Institute for Molecular Science, Okazaki 444-8585, Japan\\
$^3$Research Reactor Institute, Kyoto University, Osaka 590-0494, Japan\\
$^4$Department of Physics, Okayama University of Science, Okayama 700-0005, Japan\\
$^5$Department of Physics, Sungkyunkwan University, Suwon 440-746, South Korea\\
$^6$Nano-Materials Laboratory, National Institute for Materials Science, Tsukuba 305-0047, Japan
}

\date{\today}

\begin{abstract}
Infrared reflectivity measurements of CeSb under multi-extreme conditions (low temperatures, high pressures and high magnetic fields) were performed.
A pseudo gap structure, which originates from the magnetic band folding effect, responsible for the large enhancement in the electrical resistivity in the single-layered antiferromagnetic structure (AF-1 phase) was found at a pressure of 4~GPa and at temperatures of 35 - 50~K.
The optical spectrum of the pseudo gap changes to that of a metallic structure with increasing magnetic field strength and increasing temperature.
This change is the result of the magnetic phase transition from the AF-1 phase to other phases as a function of the magnetic field strength and temperature.
This result is the first optical observation of the formation and collapse of a pseudo gap under multi-extreme conditions.
\end{abstract}

\pacs{71.27.+a, 78.20.Ls, 75.30.Kz}
\maketitle

Ce monopnictides have many magnetic phases with complex magnetic structures at low temperatures, high pressures and high magnetic fields.~\cite{ref01}
The magnetic moments originate from the localized Ce$^{3+}4f^1$ electron.
Direct magnetic exchange interaction between Ce$4f$ electrons does not occur, but the magnetic interaction mediated by the conduction and valence electrons plays an important role.
In the case of CeSb, the interaction is known to be due to the hybridization between the Ce$4f$ and Sb$5p$ orbitals, referred to as $pf$ mixing.~\cite{ref02}

The magnetic properties and magnetic structures of CeSb have been thoroughly investigated by neutron scattering experiments.~\cite{ref03,ref04}
However, only a few studies on the modulation of the Sb$5p$ bands due to the $pf$ mixing have been performed.
In particular, at pressures of several GPa, the electrical resistivity ($\rho\sim$ several m$\Omega$cm) at around $T$ = 30~K increases by one full order over that at ambient pressure ($\rho\sim140\mu\Omega$cm).~\cite{ref05}
The origin of this increase is believed to be the strong modulation of the Sb$5p$ bands.
In this Letter, we show that the Sb$5p$ band modulation can be convincingly confirmed by optical measurements.

The magnetic phase in which the enhancement appears is the single-layered antiferromagnetic (AF-1) phase, which is not present at ambient pressure.
The magnetic structure is $+-$, where $+$ and $-$ indicate the magnetic moment direction of the ferromagnetic layer along the magnetic field.
The double-layered antiferromagnetic ($++--$, AF-1A) phase, which is different from the AF-1 phase, appears at ambient pressure and also at high pressures.
The only difference between the AF-1A and AF-1 phases is in their magnetic structures, but $\rho$ in the AF-1A phase ($\rho\leq10\mu\Omega$cm) is that of a metallic phase, which is also different from that in the AF-1 phase.
Clarification of the difference in the electronic structures of the AF-1 and AF-1A phases and their temperature and magnetic field dependencies is fundamental for understanding the changes in the Sb$5p$ band due to the different magnetic structure.

Angle-resolved photoemission~\cite{ref06,ref07} and infrared reflection spectroscopy~\cite{ref08,ref09} are employed to observe the electronic structure.
The photoemission spectroscopy cannot be performed at high pressures or in the presence of magnetic fields, in which cases infrared spectroscopy is used.
Recently, Kimura {\it et al.} reported a change in the electronic structure of CeSb due to a change in the magnetic order using infrared reflection spectroscopy and magnetic circular dichroism.~\cite{ref10,ref11}
The change in the Sb$5p$ bands due to $pf$ mixing in addition to the hybridization between the Sb$5p$ and Ce$5d$ orbitals plays an important role in the formation and stabilization of the double-layered magnetic structure.~\cite{ref12}

The electronic structure of CeSb at high pressures has been investigated using infrared spectroscopy in our previous paper.~\cite{ref13}
In the paper, the spectral changes of $R(\omega)$ due to two magnetic phase transition at $P$~=~2.5~GPa were observed.
However, the dependences of temperature and magnetic field were not performed.
In this study, to clearly investigate the electronic structure of CeSb in the AF-1 phase and its temperature and magnetic field dependencies, the higher pressure of 4.0~GPa is applied to the sample to stabilize the AF-1 phase.
And then $\sigma(\omega)$ spectra at 4.0~GPa, low temperatures and magnetic fields are investigated.
As the results, we found that the formation of the pseudo gap at the Fermi level ($E_{\rm F}$) due to the magnetic band folding effect and its collapse by temperature and magnetic field in $\sigma(\omega)$.
This is the first ever report involving infrared reflection spectroscopy under multi-extreme conditions.

Single crystalline samples of CeSb were grown by the Bridgman method using a tungsten heater crucible.~\cite{ref14}
A 0.2~$\times$~0.2~$\times$~0.05~mm$^3$ sample was set in a diamond anvil cell with Apiezon-N grease as the pressure medium and with a gold film on the sample as a reference.
The pressure was calibrated using a ruby fluorescence method.
Infrared reflectivity measurements were performed at the magneto-optical imaging station of BL43IR at SPring-8, Hyogo, Japan.~\cite{ref15}
The station is equipped with an infrared microscope, a superconducting magnet, and a cryostat for cooling samples.~\cite{ref16}
The diamond anvil cell was mounted at the sample position.
The measurement parameters include the photon energy range of 0.1 - 1~eV, temperatures of 4 - 70~K, magnetic fields of 0 - 14~T and a pressure of 4.0$\pm$0.1~GPa.
A laboratorial infrared microscopy system without a magnetic field was also used in the photon energy range of 0.065 - 1.5~eV.
To derive the $\sigma(\omega)$ spectra via the Kramers-Kronig analysis (KKA), the obtained infrared $R(\omega)$ spectra were connected to the $R(\omega)$ spectrum in the photon energy region of 1.2 - 200~eV recorded only at room temperature~\cite{ref11} and were extrapolated using $R(\omega) \propto \omega^{-4}$ above 200~eV.~\cite{ref17}
In the lower energy region, Hagen-Rubens function (HRF) ($R(\omega) = 1 - [2 \cdot \omega / (\pi \sigma_0)]^{1/2}$, where $\sigma_0$ is the direct current conductivity) was used for the metallic spectra.
In the event case that the HRF cannot be connected to the obtained $R(\omega)$ spectra, a constant extrapolation was employed.
This situation would not indicate metallic electronic structures.
A detailed analysis is shown below.

\begin{figure}[t]
\begin{center}
\includegraphics[width=6.0cm]{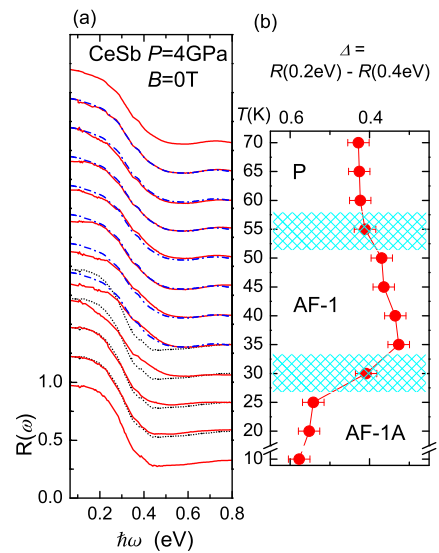}
\end{center}
\caption{
(a) Temperature dependence of the reflectivity spectrum of CeSb at $P$ = 4~GPa and $B$ = 0~T.
Successive curves are offset by 0.25 for clarity.
The spectra at 10~K and 70~K are also plotted by dotted and dot-dashed lines, respectively, in the intermediate temperature range as a guide.
(b) The difference ($\Delta$) between the reflectivity at 0.2~eV and that at 0.4~eV as a function of temperature.
The hatched areas are the phase transition temperatures.
The phase names of AF-1A, AF-1 and P follow the same notation as for neutron scattering.~\cite{ref03}
}
\label{fig1}
\end{figure}
The temperature dependence of the $R(\omega)$ spectrum at $P$ = 4~GPa and $B$ = 0~T is shown in Fig.~\ref{fig1}(a).
At $T$ = 10 and 70~K, the spectra are typically metallic because $R(\omega)$ approaches unity with decreasing photon energy.
At intermediate temperatures, the $R(\omega)$ spectrum displays a strong temperature dependence.
At 30~K, the spectrum changes drastically.
Particularly, $R(\omega)$ below 0.2~eV does not approach unity and cannot be fitted by HRF.
This means that the metallic character is suppressed at intermediate temperatures.

To clarify the temperature dependence of $R(\omega)$, the differential intensity ($\Delta$) between $R(\hbar\omega = 0.2~{\rm eV})$ and $R(\hbar\omega = 0.4~{\rm eV})$ is plotted as a function of temperature in Fig.~\ref{fig1}(b).
This figure indicates that the temperature dependence of $\Delta$ is very weak below 20~K and above 60~K.
A large jump appears at 30~K and $\Delta$ slightly increases with increasing temperature between 35 and 50~K.
At around 50~K, $\Delta$ smoothly connects to the constant value above 60~K.
The two characteristic temperatures shown by the hatched area is in good agreement with the magnetic phase transition temperatures at 4~GPa determined by other methods.~\cite{ref04,ref05}
Based on this figure, it is clear the electronic structures in the AF-1A and P phases are almost constant.
At the transition temperature from the AF-1A to AF-1 phases, the electronic structure drastically changes.
In the AF-1 phase, the electronic structure gradually changes with increasing temperature, eventually reaching the value of that in the P phase.
This is consistent with the behavior of the intensity at the sublattice peak $[1,1,0]$ indicating the presence of the AF-1 phase magnetic structure detected by neutron scattering.~\cite{ref03}
The magnetic scattering intensity at the sublattice peak as a function of temperature illustrates the second-order nature of the transition from the P to AF-1 phases and the first-order one from the AF-1 to AF-1A phases.
This behavior is consistent with the temperature dependence of $\Delta$.
This directly provides that the magnetic structure strongly couples to the Sb$5p$ electronic structure.
This same result has been obtained in CeBi at ambient pressure.~\cite{ref18}
CeBi has an AF-1 phase even at ambient pressure.
In the CeBi case, the $pf$ mixing intensity increases with decreasing temperature.
CeSb at $P$ = 4~GPa gives the same result, indicating the temperature dependent $pf$ mixing intensity is responsible for the change in $\Delta$.

\begin{figure}[t]
\begin{center}
\includegraphics[width=5.5cm]{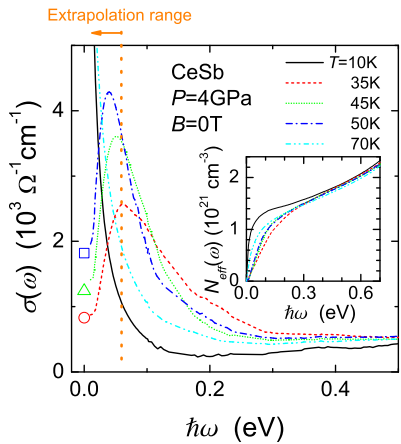}
\end{center}
\caption{
Temperature dependence of the optical conductivity spectrum ($\sigma(\omega)$) of CeSb at $P$ = 4~GPa and $B$ = 0~T.
The direct current conductivities at $T$ = 35~K ($\bigcirc$), 45~K ($\triangle$) and 50~K ($\Box$) are also plotted.
Since the measured energy range was above 0.65~eV, the $R(\omega)$ in the energy range below 0.65~eV was extrapolated using Hagen-Ruben's function at 10 and 70~K or the modified Hagen-Ruben's function at 35, 45 and 50~K for the Kramers-Kronig analysis.
See the text for detail.
(Inset) The effective electron number ($N_{eff}(\omega)$) derived from $\sigma(\omega)$.
}
\label{fig2}
\end{figure}
To clarify the change in the electronic structure at $P$ = 4~GPa and $B$ = 0~T as a function of temperature, $\sigma(\omega)$ derived from the KKA of $R(\omega)$ is shown in Fig.~\ref{fig2}.
To perform the KKA, the HRF derived from $\sigma_0$ must smoothly connect to $R(\omega)$ at the temperatures below 25~K and above 55~K.
The obtained $\sigma(\omega)$ spectra at 10 and 70~K shown in Fig.~\ref{fig2} look like normal metallic ones.
However, the HRF derived from $\sigma_0$ does not connect to $R(\omega)$ at the temperatures between 30 and 50~K.
This means that the measured $R(\omega)$ spectra are not the tail of the metallic Drude function but a part of low-energy interband transitions like in insulators.
So, $R(\omega)$ at $\hbar \omega$ = 0.065~eV is extrapolated to the lower energy side as a constant and is connected to the HRF derived from $\sigma_0$.
The absolute value of $R(\omega)$ must contain the physical properties at around this energy.
Therefore, the $\sigma(\omega)$ spectrum in the extrapolated region has a qualitative meaning.
To evaluate this quantitatively, these measurements in the lower energy region under the same multi-extreme conditions should be performed, which is impossible at present.
The $\sigma(\omega)$ spectra derived from the KKA using this extrapolation method are plotted in Fig.~\ref{fig2}.
The $\sigma(\omega)$ spectrum at 35~K has a peak with a peak energy of 0.065~eV.
The peak energy, which originates from both of the absolute value of $R(\omega)$ and $\sigma_0$, shifts to the lower energy side with increasing temperature.
In CeBi, a similar peak appears in the AF-1 phase.~\cite{ref18}
The peak shifts to the low energy side with increasing temperature in the same way as that of CeSb.
However, CeBi differs in that it has a metallic $\sigma(\omega)$ spectrum and also a metallic $\sigma_0$ value even in the AF-1 phase.
This indicates that the AF-1 phases in both CeSb and CeBi have a gap shape electronic structure in common.
However, the gap in CeBi is distant from $E_{\rm F}$ and that of CeSb is just on $E_{\rm F}$.
This directly accounts for the different behavior in $\rho$ between these materials.

To investigate the origin of the pseudo gap, the effective electron number ($N_{eff}(\omega)=\frac{2 m_0}{\pi e^2} \int_0^\omega \sigma(\omega_0) d\omega_0$) is plotted in the inset of Fig.~\ref{fig2}.
Here $m_0$ means free electron mass.
Based on the figure, the temperature dependence occurs below 0.4~eV.
This means that only the electronic structure 0.4~eV below $E_{\rm F}$ depends on temperature.
The energy of 0.4~eV is similar to the $pf$ mixing energy ($E(pf\sigma)=$0.35~eV, $E(pf\pi)=$-0.245~eV) of CeSb.~\cite{ref02, ref12}
This indicates that the Sb$5p$ band modified by the Ce$4f$ spin structure due to the $pf$ mixing is the origin of the pseudo gap.
The enhancement in the electrical resistivity originates from the decrease in the density of states on $E_{\rm F}$ through the creation of the pseudo gap structure.
A similar pseudo gap structure is also observed in cases of pairing due to a charge/spin density wave~\cite{ref19}, of a lattice distortion due to the Jahn-Teller effect~\cite{ref20} etc.
In the present case, no pairing effect was observed.
On the other hand, a very small lattice contraction appears in the magnetically ordered states in the form of a change in the Ce$4f$ ground state from $\Gamma_7$ with a large charge distribution in the P phase to $\Gamma_8$ with a small distribution in the AF-1 phase.~\cite{ref21,ref22}
However, if the lattice distortion would be the origin of the pseudo gap, the gap structure would remain in the AF-1A phase, as the $\Gamma_8$ ground state is common even in the AF-1A phase.
Since the difference between the AF-1 and AF-1A phases is only in the alignment of the magnetic moments, the origin of the pseudo gap must be the periodicity of the Ce$4f$ magnetic moments contained in the Sb$5p$ band through the $pf$ mixing.
This is referred to as the magnetic band folding effect.

\begin{figure}[t]
\begin{center}
\includegraphics[width=6.0cm]{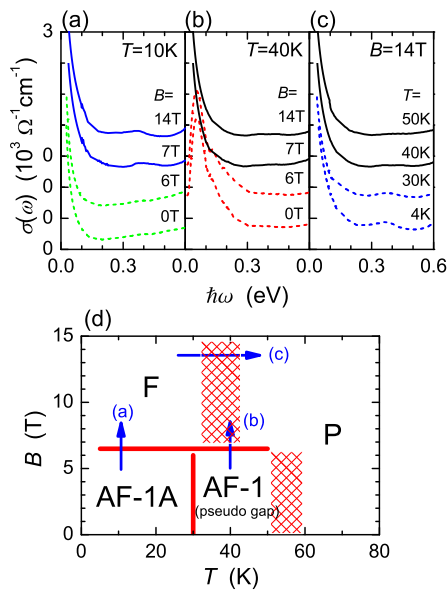}
\end{center}
\caption{
Optical conductivity ($\sigma(\omega)$) spectra of CeSb at $P$ = 4~GPa.
(a) Magnetic field dependence at $T$ = 10~K.
(b) Magnetic field dependence at $T$ = 40~K.
(c) Temperature dependence at $B$ = 14~T.
(d) Magnetic field - temperature $(B, T)$ phase diagram of CeSb at $P$ = 4~GPa derived from $\sigma(\omega)$ spectra under multi-extreme conditions.
The first-order and second-order like spectral changes are shown by lines and hatchings, respectively.
The arrows indicate the scans (a) - (c).
}
\label{fig3}
\end{figure}
Next, we show an example of a magnetic-field-induced nonmetal-metal transition under a high pressure.
Figure~\ref{fig3} indicates the magnetic field dependence of $\sigma(\omega)$.
In Fig.~\ref{fig3}(a), the $\sigma(\omega)$ spectrum at $B$ = 0~T is equal to that at 6~T.
The spectrum at 7~T is different from that at 6~T and is equal to that at 14~T.
This indicates that the region below 6~T is all the same phase and that above 7~T is a different one.
The phase transition magnetic field is 6.5 $\pm$ 0.5~T.
The data in Fig.~\ref{fig3}(b) indicate the same situation as that in Fig.~\ref{fig3}(a).
However, the spectrum at $T$ = 40~K and $B$ = 0~T is different from that at 10~K and 0~T because these magnetic phases are different, as shown in Fig.~\ref{fig1}.
Since the $\sigma(\omega)$ spectrum at $T$ = 10~K and $B$ = 14~T is the same as that at ambient pressure, 6.5~K and 6~T for the ferromagnetic (F) phase,~\cite{ref10} the condition of $P$ = 4~GPa, $T$ = 10~K and $B$ = 14~T is also in the F phase.
The transition magnetic field from the AF-1A to F phases shifts from 4.4~T at ambient pressure to 6.5~T at 4~GPa.
The reason for this shift is that the $pf$ mixing increases and stabilizes the double-layered magnetic structure with a decrease in the distance between the Ce and Sb atoms as a result of the applied pressure.
At 40~K in Fig.~\ref{fig3}(b), the pseudo gap structure appears below $B$ = 6~T.
The gap structure disappears at 7~T across the transition magnetic field of 6.5~$\pm$~0.5~T, and the $\sigma(\omega)$ spectrum changes to a metallic one.
This is an example of a magnetic-field-induced nonmetal-metal transition under a high pressure.

At 0~T, the $\sigma(\omega)$ spectrum drastically changes with temperature because of the magnetic phase transition, as shown in Fig.~\ref{fig2}.
The same spectral transition at the same transition temperature was observed below 6~T (not shown).
Above 7~T, a different spectral change was observed, in which the $\sigma(\omega)$ spectrum displayed a more typical temperature dependence, shown in Fig.~\ref{fig3}(c).
This temperature dependence indicates that the F phase below 30~K gradually changes to a different phase above 50~K.
Since the $\sigma(\omega)$ spectrum above 50~K is the same as that at 70~K and 0~T, as shown in Fig.~\ref{fig2}, the higher temperature phase must be the P phase.
Based on this, these results indicate the F phase gradually changes to the P phase with increasing temperature in the magnetic field range above 7~T.

The magnetic field - temperature $(B, T)$ phase diagram at 4~GPa resulting from changes in the $\sigma(\omega)$ spectrum is shown in Fig.~\ref{fig3}(d).
In the figure, the solid lines and the hatchings are the first-order-like and second-order-like spectral transitions, respectively.
The temperature and magnetic field region of the AF-1A phase expands at 4~GPa in comparison with that at ambient pressure.
This is the plausible result of the $pf$ mixing increasing due to the applied pressure.
On the other hand, the phase diagram at 4~GPa becomes simpler than that at ambient pressure. 
In particular, the complex magnetic structure, for example, in AFP$x$ ($x~=~1~\sim~6$) phases at ambient pressure~\cite{ref01} disappears at 4~GPa.
At ambient pressure, since the $pf$ mixing competes with other magnetic interactions and crystal field splitting, such complex magnetic phases and structures appear.
With increasing pressure, the $pf$ mixing increases and then dominates among these interactions.
The increase of the $pf$ mixing also makes the simpler magnetic phase diagram at 4~GPa.

In summary, infrared reflection spectroscopy on CeSb at low temperatures, high pressures and high magnetic fields was performed.
The pseudo gap structure was clearly observed in the AF-1 phase at 4~GPa, as well as the collapse of the gap with increasing magnetic field.
The pseudo gap originates from the magnetic band folding effect.
CeSb has the necessary electronic structure to create the pseudo gap in the AF-1 phase.
CeSb in the AF-1 phase is the model system of the magnetic band folding.
This paper indicates the first optical observation of the magnetic field induced nonmetal-metal phase transition at high pressures.
The magnetic field - temperature phase diagram of CeSb at 4~GPa was also determined.

We would like to thank to Drs. Kaneta and Sichelschmidt for valuable discussions and also thank SPring-8 BL43IR staff members for their technical support.
This work was partially supported by a Grant-in-Aid for Young Scientists (A) (Grant No. 14702011) from MEXT of Japan.
The experiments were performed with the approval of the Japan Synchrotron Radiation Research Institute (Proposal Nos. 2003A0076-NS1-np, 2004A0231-NSa-np and 2004B0545-NSa-np).


\end{document}